\documentclass[conference]{IEEEtran}
\usepackage{amsmath,amssymb,amsfonts}
\usepackage{graphicx}
\usepackage{filecontents,lipsum}
\usepackage{cite}
\usepackage[labelfont=bf]{caption}
\usepackage{blindtext}
\usepackage{float}
\usepackage[utf8]{inputenc}

\title{Waveform Phasicity Prediction from Arterial Sounds through Spectrogram Analysis using Convolutional Neural Networks for Limb Perfusion Assessment}

\author{\IEEEauthorblockN{Adrit Rao}
\IEEEauthorblockA{\textit{Greene Middle School} \\
Palo Alto, CA, USA \\
adrit.rao@gmail.com}
\and
\IEEEauthorblockN{Kevin Battenfield, RVT}
\IEEEauthorblockA{\textit{Stanford University} \\
Stanford, CA, USA\\
kbatt@stanford.edu \\
}
\and
\IEEEauthorblockN{Oliver Aalami, MD}
\IEEEauthorblockA{\textit{Stanford University} \\
Stanford, CA, USA\\
aalami@stanford.edu}
}

\begin{document}

\maketitle 
\begin{abstract}

Peripheral Arterial Disease (PAD) is a common form of arterial occlusive disease that is challenging to evaluate at the point-of-care. Hand-held dopplers are the most ubiquitous device used to evaluate circulation and allows providers to audibly "listen" to the blood flow. Providers use the audible feedback to subjectively assess whether the sound characteristics are consistent with Monophasic, Biphasic, or Triphasic waveforms. Subjective assessment of doppler sounds raises suspicion of PAD and leads to further testing, often delaying definitive treatment. Misdiagnoses are also possible with subjective interpretation of doppler waveforms. This paper presents a Deep Learning system that has the ability to predict waveform phasicity through analysis of hand-held doppler sounds. We collected 268 four-second recordings on an iPhone taken during a formal vascular lab study in patients with cardiovascular disease. Our end-to-end system works by converting input sound into a spectrogram which visually represents frequency changes in temporal patterns. This conversion enables visual differentiation between the phasicity classes. With these changes present, a custom trained Convolutional Neural Network (CNN) is used for prediction through learned feature extraction. The performance of the model was evaluated via calculation of the F1 score and accuracy metrics. The system received an F1 score of 90.57\% and an accuracy of 96.23\%. Our Deep Learning system is not computationally expensive and has the ability for integration within several applications. When used in a clinic, this system has the capability of preventing misdiagnosis and gives practitioners a second opinion that can be useful in the evaluation of PAD.

\end{abstract}


\section{Introduction}

Peripheral Arterial Disease (PAD) \cite{ouriel2001peripheral} is the narrowing or blockage of arteries supplying the lower extremities due to atherosclerosis (Figure 1). According to the Center for Disease Control (CDC) over 6.5 million people who are 40 years or older have PAD in the United States alone \cite{cdc}. If not diagnosed and treated early, PAD can lead to limb loss. After clinical examination for a pedal pulse, the most common point-of-care tool utilized for evaluation of PAD is a low-cost hand-held doppler \cite{tehan2015use}. When pointed at the direction of blood flow of a target artery, this device allows the provider to audibly "listen" to the circulation. Providers are then expected to subjectively assess whether the sound characteristics are consistent with Monophasic, Biphasic, or Triphasic waveforms to diagnose the severity of PAD. This is a technically challenging procedure as waveforms can be difficult to predict. Thus when manually assessing these sounds mistakes can be made leading to misdiagnoses. With any question in sound interpretation, practitioners will often refer patients for further testing to a formal vascular ultrasound laboratory which can lead to delays in definitive treatment. Within the vascular laboratory, the Ankle-Brachial Index (ABI) is calculated and/or direct arterial ultrasound imaging can be performed to diagnose the severity of PAD. For this, large and expensive devices are utilized.

\begin{figure}[h!]
    \centering
    \includegraphics[scale=0.45]{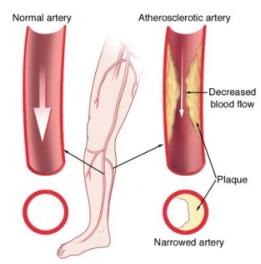}
    \caption{CDC PAD Artery Illustration \cite{cdc}}
    \captionsetup{labelfont=bf}
    \label{fig:architecture}
\end{figure}

Taking into account the subjectivity when assessing arterial sounds and predicting waveform phasicity, there is a clear need for an effective method for practitioners to cross-check their prediction and increase their confidence in order to provide more timely treatment of PAD. This paper presents a novel Deep Learning system that has the ability to predict waveform phasicity through analysis of hand-held doppler sounds. Deep Learning is becoming increasingly popular in the medical field for its ability to mimic the human brain structure, learn from data and make clinically useful predictions \cite{lee2017deep}.

With current computational resources, Deep Learning is being widely utilized for multiple tasks. Most clinical implementations of Deep Learning solutions are centered around image classification and object detection for medical imaging purposes \cite{lee2017deep}. Our method employs similar Deep Learning techniques but for the classification of arterial sounds. This classification is made possible by firstly converting input sound into a spectrogram image. This is done to visualize features and convert frequency and amplitude variance into visual temporal changes. With this conversion, visual differentiation is present between the phasicity classes opening up the ability for classification through a Deep Learning approach. A custom Convolutional Neural Network (CNN) \cite{krizhevsky2012imagenet} trained on spectrogram images spanning the three waveform phasicity classes is used for prediction. The CNN model revealed high accuracy rates when tested on new data. This system can enhance the capabilities of low-cost hand-held dopplers and aid practitioners at the point-of-care leading to more timely clinical decisions. Figure 2 shows the three phasicity waveforms. The Triphasic waveform consists of three phases, Biphasic has two phases and the Monophasic waveform has one phase. The spectrogram visual differentiation is similar to the waveforms visual differentiation.

\begin{figure}[h!]
    \centering
    \includegraphics[scale=0.25]{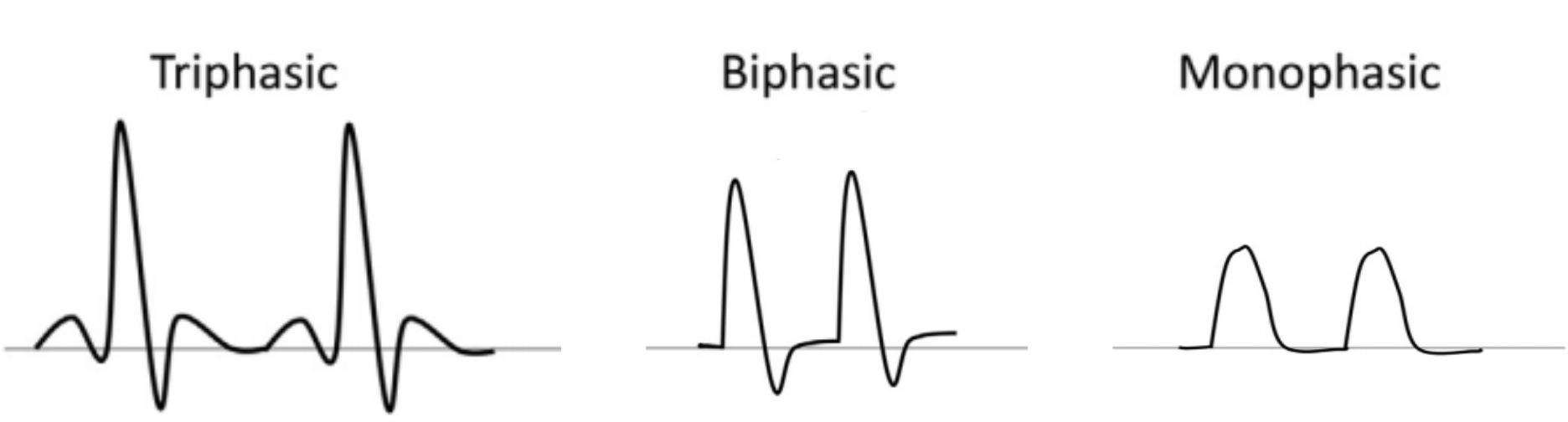}
    \caption{Waveform Phasicity Comparison}
    \captionsetup{labelfont=bf}
    \label{fig:architecture}
\end{figure}

\section{Related Work}

A large amount of research has been conducted around the usage of Deep Learning, Machine Learning (ML) and Computer Vision (CV) based techniques and algorithms for the interpretation of data to classify cardiovascular conditions and diseases. We have highlighted a paper that utilizes a sound classification system for the classification of heart sounds (eg., murmurs). Li et al. proposed the use of a custom trained CNN for the classification of heart sounds as either abnormal or normal \cite{li2020classification}. This research utilized the PhysioNet cardiology dataset \cite{costa2003physionet}. They used various signal processing techniques for the plotting of Mel-frequency Cepstral Coefficients (MFCC). They utilized a custom CNN to analyze the MFCC's and make the binary classification. The proposed algorithm received around a 72.1\% to 86.8\% accuracy rate in classification.

Our method differentiates from Li et al. in two main factors. Firstly, our method targets peripheral arteries affected by PAD and not the heart itself. Thus our research required different means for both data collection, generation, and model development. Additionally, in our opinion, making a binary classification such as normal or abnormal likely does not provide a detailed enough result for practitioners to interpret and utilize. Our system predicts waveform phasicity, a widely utilized metric for the diagnosis of PAD and other cardiovascular diseases. With a predicted phasicity, practitioners can differentiate between mild, moderate, and severe arterial disease. This can help in making a more robust diagnosis. Our system has also achieved an accuracy rate of over 90\%.

\section{Methods}

The following order and methodology has been used to develop and validate our full end-to-end sound classification system. Firstly, audio data was collected and labeled within a clinical setting by a vascular technologist during formal ABI examinations. After the collection of data, spectrogram images were generated by inputting all collected sound files through a signal processing conversion method. This converted frequency changes into temporal changes. As the dataset was not very large in size, artificial data augmentations were added to increase dataset size and possible learned features. After data preparation, a Deep Learning model was developed and constructed through a Transfer Learning \cite{tl} approach enabling high accuracy rates and less training time on lower amounts of data. Lastly, the model was validated with a separate set of doppler audio data through calculation of the F1 score and accuracy metrics. Figure 3 depicts the fully constructed sound classification system flow. The end-to-end system works by converting input sound into a spectrogram image and then sending the image through a custom trained CNN which predicts the waveform phasicity through learned visual features.

\begin{figure}[h!]
    \centering
    \includegraphics[scale=0.28]{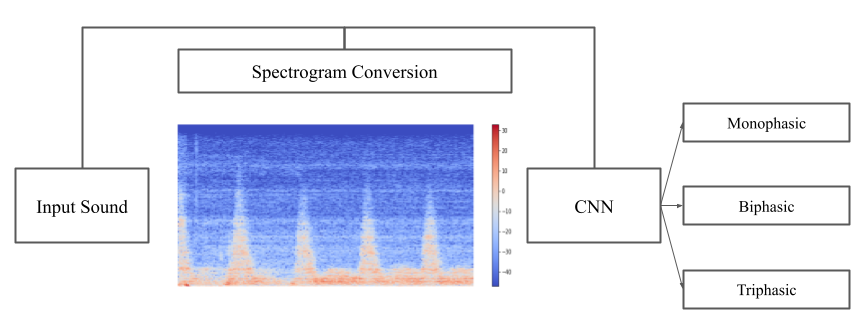}
    \caption{Full Sound Classification System Flow}
    \captionsetup{labelfont=bf}
    \label{fig:architecture}
\end{figure}

As follows are each of the development stages consisting of a detailed explanation which walks through the exact methodology as well as well as specific tools and formulas used for implementation of the full sound classification system. 

\subsection{Data Collection}

Hand-held doppler sounds were generated using the Parks Flo-Lab 2100 doppler machine pencil doppler \cite{parks}. Four-second recordings were taken during formal vascular lab studies in patients being evaluated for PAD. An iOS application was developed to efficiently and securely create and label the recordings through an iPhone within the clinic. The vascular technologist performing the study labeled the waveforms using the Parks Flo-Lab digital waveform display as a robust reference to maintain accuracy and data consistency. The iOS iPhone application was developed within Apple's Xcode Integrated Development Environment (IDE) with the Swift 5 programming language. The app was linked to a Cloud database via Google's Firebase \cite{google} allowing for real-time and secure data transmission. Within the Firebase system, a real-time Database \cite{google} was created for storage of input values such as waveform phasicity and arterial position, and a Storage bucket was created for storage of sound files. Phasicity and artery names were also appended to the sound file titles for fast referencing of data for cross-checking.

After data was collected within the database, the Python 3 programming language and the Pyrebase third-party data extraction library was used to create a data downloading script utilized for downloading all files from the real-time database and storage bucket at once. Shown in Figure 3, is the dataset distribution after all data was extracted from the database. A total of 268 sound files were collected. The dataset was split into an 80\% training and a 20\% testing standard split. This was to ensure data was available after model training for validation/testing of the system.

\begin{figure}[H]
    \centering
    \includegraphics[scale=0.5]{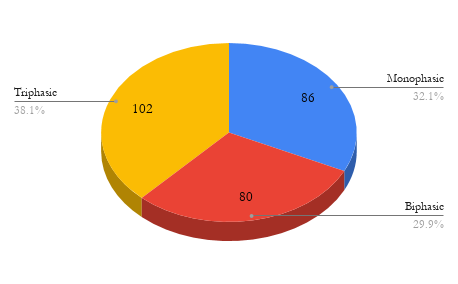}
    \caption{Dataset Distribution Chart}
    \captionsetup{labelfont=bf}
    \label{fig:architecture}
\end{figure}

Diverse data with background interference and noise was intentionally included in the final dataset. This was to ensure that in later stages, the model would potentially develop resilience to these factors as they will likely be present in a clinical setting due to noise from other machines and tools.

\subsection{Data Generation}

All sound files were converted into spectrogram images that represent frequency changes visually over the four-second time span. The color variance in the spectrogram represents amplitude changes. These changes clearly visually differentiate the Monophasic, Biphasic, and Triphasic classes. Shown in Figure 4, are spectrograms representing each of the three phasicity classes. The Monophasic class does not have a lot of definition in the amplitude variance but in the Biphasic and Triphasic classes the amplitude variance increases showing clear visual class differentiation. This differentiation is a representation of the increased blood flow and pulsatility present in the Biphasic and Triphasic waveform phasicity classes. Also, note that interference, as well as background noise, is present in the selected spectrogram samples but differentiation remains.

\begin{figure}[h!]
    \centering
    \includegraphics[scale=0.3]{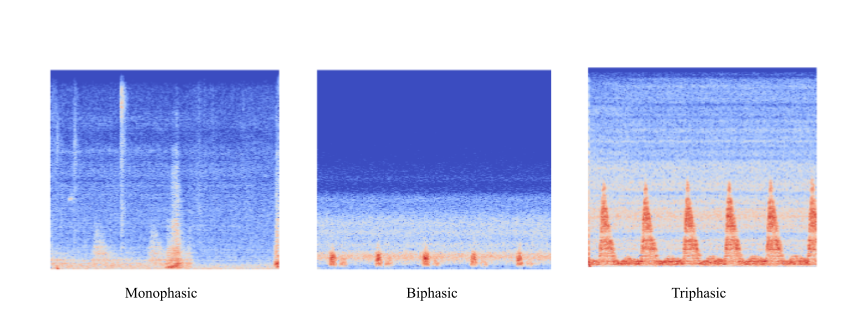}
    \caption{Spectrogram Phasicity Comparison}
    \captionsetup{labelfont=bf}
    \label{fig:architecture}
\end{figure}

Sound file to spectrogram conversion was implemented in the Python 3 programming language with the Librosa library \cite{mcfee2015librosa}. Librosa is a popular library for sound and signal processing as it holds a lot of useful features within a minimal and simplified codebase. Spectrogram generation was done within a loop function. This function converted input sound into a spectrogram image by sending it through a Short-time Fast Fourier Transform (STFT) \cite{allen1977short}. The STFT is a widely used and standard conversion method for spectrogram generation. At a high level, this type of Fast Fourier Transform (FFT) \cite{brigham1988fast} divides a time signal into shorter equal segments/windows, runs a standard FFT, and plots the spectrogram in the time-frequency domain. Shown in Equation 1 is the mathematical formula for the STFT. In this formula, the signal is represented as \(x[n]\) and the window as \(w[n]\). The standard FFT converts the input signal from its original domain into a frequency domain representation. Doing this within the STFT method allows for spectral plotting. After the generation of spectrograms, all images were converted into a [224, 224] constant size as this is required by the specific model developed in later stages.

\begin{equation}
\mathbf{STFT}\{x[n]\}(m, \omega) \equiv X(m, \omega)=\sum_{n=-\infty}^{\infty} x[n] w[n-m] e^{-j \omega m}
\end{equation}

\subsection{Data Augmentation}

After all spectrogram images were generated, the dataset was artificially augmented to increase the size and possible knowledge gained by the model in later stages. The Keras Python library was used for adding data augmentations as sequential layers. Keras is a widely used framework for Neural Network construction as well as image pre-processing. One simple data augmentation was added due to the system taking in clean spectral images in the end-stage. Randomized zoom (20\%) was chosen as the primary augmentation for this dataset. Randomized zoom maintains valuable features as well as orientation. Figure 5 shows an example Triphasic spectrogram image with randomized zoom applied. Amplitude variance is still visible in the image. This can possibly be valuable.

\begin{figure}[H]
    \centering
    \includegraphics[scale=0.18]{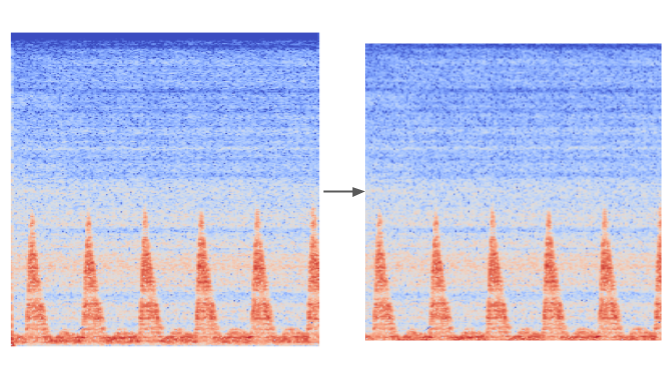}
    \caption{Randomized Zoom Data Augmentation}
    \captionsetup{labelfont=bf}
    \label{fig:architecture}
\end{figure}

\subsection{Model Development}

Deep Learning is emerging as a powerful technology that is being used widely in the medical field to aid practitioners. Deep Learning gives computers the ability to learn from data and make clinically useful predictions by mimicking the human brain's neural structure and enabling the development of Artificial Neural Networks (ANNs). CNNs, a form of Deep Neural Networks (DNNs) have revealed high accuracy rates when analyzing visual imagery due to their unique capability of extracting features through convolutions. As with many medical projects, our dataset was small, and training a CNN from scratch would not be ideal and high accuracy rates would not be achievable. To compensate for the lack of data, a Transfer Learning approach was leveraged. Transfer Learning is a popular Deep Learning methodology that allows for the retraining/relearning of existing models. These models usually have deep architectures consisting of multiple layers and have been trained on large amounts of images. The prior knowledge gained from the model helps with the differentiation of new classes thus gaining the ability to achieve high accuracy rates on minimal amounts of data. This approach also enables faster training time (fewer epochs/iterations) compared to developing a Neural Network from scratch with no prior knowledge.

Inception V3 \cite{szegedy2016rethinking} is a popular CNN model architecture that has received fairly high accuracy for image classification problems (ImageNet). Inception V3 builds upon previous Inception CNN versions due to its ability to reduce computation and enable deep networks through dimensionality reduction techniques. Inception V3 is a vast network architecture consisting of multiple layers of convolution, pooling, concat, dropout, etc. The Inception V3 model utilized for this project has gained prior knowledge through being trained on around a million images from the ImageNet dataset/competition \cite{deng2009imagenet}. Because of this, the model has gained weights that can help with the differentiation of newly trained classes. To accommodate the new spectrogram image dataset, two layers were added to the end of the Inception V3 architecture. The modified Inception architecture consists of both a Flatten layer and a Dense layer for final probability class prediction. The Flatten layer converts the output of the original Inception V3 architecture into a one-dimensional representation. The Dense layer is used for final probability class prediction through the softmax activation function (Equation 2). Softmax is commonly used in Neural Networks for multiple class predictions. Our softmax function contained three nodes for the phasicity classes. 

\begin{equation}
\mathbf{\sigma(\vec{z})_{i}}=\frac{e^{z_{i}}}{\sum_{j=1}^{K} e^{z_{j}}}
\end{equation}

The CNN model was constructed and implemented sequentially in Keras. Compilation and training were done with Tensorflow \cite{abadi2016tensorflow}. The Google Colab Pro Integrated Development Environment (IDE) \cite{bisong2019google} was leveraged for Cloud development allowing for fast model training with large amounts of available compute. The Python 3 programming language was utilized. Within Colab, the Tesla P100 GPU \cite{nvidia} was leveraged. The modified Inception V3 architecture is shown in Figure 7. The full Inception V3 model was downloaded from Google Storage with the ImageNet weights initialized.

\begin{figure}[h!]
    \centering
    \includegraphics[scale=0.4]{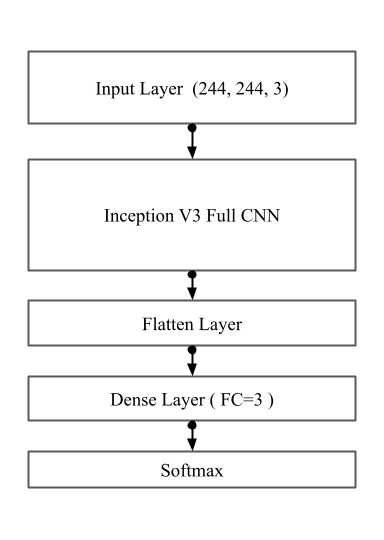}
    \caption{Modified Inception V3 Model Architecture}
    \captionsetup{labelfont=bf}
    \label{fig:architecture}
\end{figure}

\subsection{Model Training}

The model was compiled first and then trained across 10 epochs using Tensorflow. When training a Neural Network from scratch, epochs can be up to 100 but in our case due to the usage of Transfer Learning, 10 epochs was sufficient for gaining high accuracy. The model was fed both training and testing data. For compilation, the Adam optimizer \cite{kingma2014adam} was utilized. Sparse Categorical-cross entropy was the loss function chosen. At each epoch in model training F1 score and accuracy, metrics were calculated as shown in Equation 4 and Equation 5. Additionally, the model's training and testing accuracies were graphed over the 10 epochs along with loss. This data when interpreted can show us if the model is learning from the data properly and if overfitting is present.

\medskip

\begin{equation}
\mathbf{F1} = \frac{2*Precision*Recall}{Precision+Recall} = \frac{TP}{TP+\frac{1}{2}(FP+FN)}
\end{equation}

\begin{equation}
\mathbf{Accuracy} = \frac{TP+TN}{TP+TN+FP+FN}
\end{equation}

\medskip

\section{Results and Discussion}

After the system was fully developed, validation was done through calculation of the F1 score and accuracy metrics. The F1 score is a standard validation metric that takes into account both precision and recall. It is widely used in the validation of Deep Learning and ML systems. Accuracy was also calculated and used in the validation of the system. Accuracy is the percentage of predictions that the model made correctly through all of the testing data. Our system received a validation accuracy of 96.23\% and an F1 score of 90.57\%. As these metrics are over 90\%, it clearly shows that the model is highly accurate in predicting phasicity through sound and that the proposed sound classification system is accurate.

To further validate the system and its learning rate, graphs created during model training were interpreted. These graphs represent training and validation accuracy and loss over the 10 training epochs. As the loss is consistently decreasing through the iterations consistently, overfitting is not present and the model has a good learning rate. The accuracy graph consistently increases without any dips indicating that the model is learning at a good rate. Additionally, the validation accuracy as well as training accuracy are close together. Figure 9 shows the loss and accuracy graphs over the 10 epochs.

Similar to most medical studies, data is limited and we could have benefited from having more data from multiple sites to help avoid any potential biases intrinsic to the patient population characteristics. FFT analysis to classify doppler signals has been attempted but never for doppler sounds of the tibial vessels \cite{serhatlioglu2003classification}. The power of such analysis is particularly evident in the setting of a calcified tibial vessel which does not allow for accurate ankle-brachial index measurement. 

\begin{figure}[h!]
    \centering
    \includegraphics[scale=0.35]{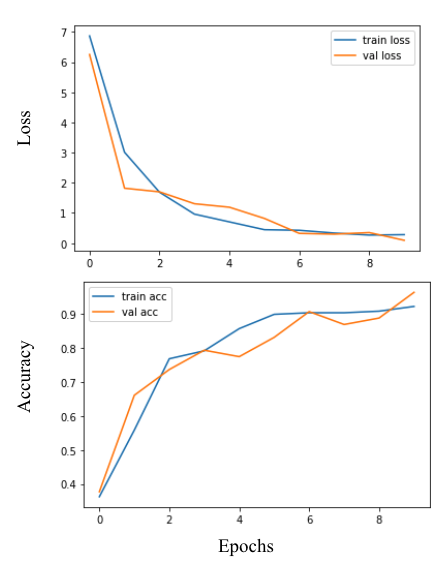}
    \caption{Model Accuracy and Loss Graphs}
    \captionsetup{labelfont=bf}
    \label{fig:architecture}
\end{figure}

\section{Conclusion}

We have successfully developed an end-to-end sound classification system that can predict arterial doppler waveform phasicity with high accuracy. Vascular specialty care is limited as are vascular labs to formally rule out PAD. This model and method can be implemented in primary care, podiatry, and wound care clinics to provide a more objective peripheral vascular assessment and assist in real-time treatment decisions and avoid delays in care. Our next steps are to study the prediction of ABIs and to integrate this model into a low-cost mobile application and hardware platform which can be used at the point-of-care ubiquitously.

\bibliographystyle{IEEEtran}
\bibliography{bib}

\end{document}